# Arabic Text Recognition in Video Sequences

Mohamed Ben Halima, Hichem Karray and Adel M. Alimi
REGIM: REsearch Group on Intelligent Machines, University of Sfax, National School of Engineers
(ENIS), BP 1173, Sfax, 3038, Tunisia
*{mohamed.benhlima, hichem.karray and adel.alimi}@ieee.org*

## Abstract

In this paper, we propose a robust approach for text extraction and recognition from Arabic news video sequence. The text included in video sequences is an important needful for indexing and searching system. However, this text is difficult to detect and recognize because of the variability of its size, their low resolution characters and the complexity of the backgrounds. To solve these problems, we propose a system performing in two main tasks: extraction and recognition of text. Our system is tested on a varied database composed of different Arabic news programs and the obtained results are encouraging and show the merits of our approach.

## 1. Introduction

Extraction and recognition of text embedded in video sequences can aid in video content analysis and understanding, while text can offer concise and direct description of the stories presented in the videos. In news broadcast, the superimposed captions usually present the involved person's name and the summary of the news event. Therefore, the recognized text can become a part of index in a video retrieval system.

Works on text extraction may be generally grouped into four categories [1]: First category is the connected component methods which detect text by extracting the connected components of monotone colours that obey certain size, shape, and spatial alignment constraints. The second is the texture methods which treat the text region as a special type of texture and employ conventional texture classification method to extract text. The third is the edge detection methods which have been increasingly used for caption extraction due to the rich edge concentration in characters [3]. Finally, the correlation based methods which use any kind of correlation in order to decide if a pixel belongs to a character or not.

News programs are particular audio-visual documents they are generally formed by a set of semantically independent stories. For this reason before starting the extraction of the textual information from the news programs, segmentation into stories will be done. The following figure shows an example of image taken from video broadcast of Aljazeera, Tunis 7, France 24 and Alarabia.

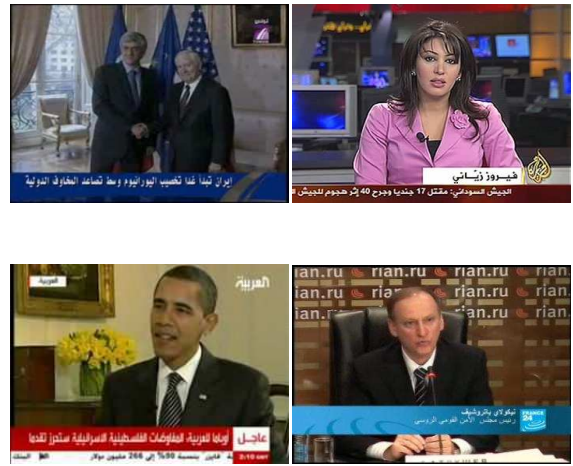

Figure1 - Examples of text in video frames

Video Text Recognition is more difficult than many other OCR applications (e.g., reading printed matter) because of degradation, such as background noise, and deformation, like the variation in fonts.

Several methods have been proposed for text recognition in videos. However, most of them are not addressed the problem of the Arabic script.

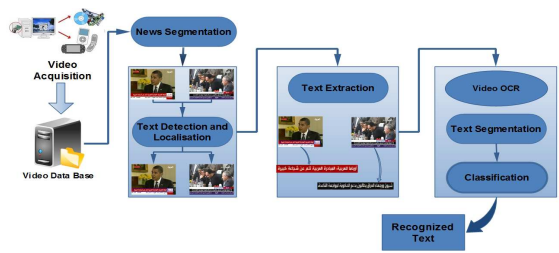

Figure 2: Global Overview of our System

In this paper, we present a new system for text extraction and recognition from Arabic news videos. Our system integrates text detection, tracking on multiple frames localisation and text recognition.

The rest of the paper will be organized as follows; in section 2, we discuss works related to news segmentation. Section 3; describe the proposed approach for text extraction. In section 4, we present our approach for text recognition. . The experiments are presented in Section 5 and conclusion in section.6.

## 2. News segmentation

Stories segmentation is an essential step in any work done on video news sequences. The story in every news program is the most important semantic unit. Story segmentation is an active research field with a lot of categories of works. However, the majority of proposed works are based on detecting anchor shots. Indeed, the anchor shot is the only repetitive shot in a news program.

In [2] [3], we use our previous work in which news program is segmented by detecting and classifying faces to find group of anchor shots. It is based on the assumption that the anchors faces are the only repetitive face throughout the entire program.

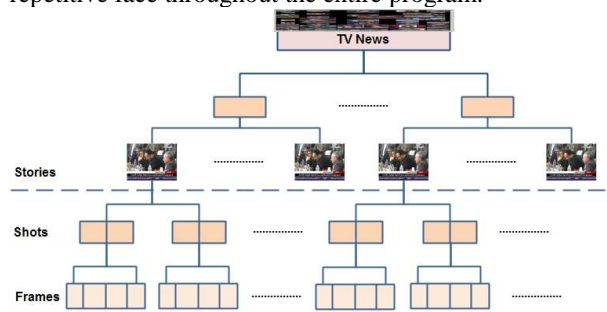

Figure 3: Structure of a news program

## 3. Text extraction

After segmenting news program into stories, we extract from every story text transcriptions. Works on text extraction may be generally grouped into four categories: connected component methods [4][5], texture classification methods [6][7], edge detection methods [8][9][10][11][12], and correlation based methods [13][14].

Generally, text embedded in video sequences can be classified into two categories: Text that is part of the scene and artificial text. Scene text (such as a sign outside a place of business or placards in front of conference participants) is part of image and usually does not represent information about image content, but artificial text (such as superimposed captions in broadcast news programs and other commercially produced videos) is laid over the image in a later stage. Artificial text is often good key for successful indexing and retrieval of videos. In news broadcast, overlaid text presents the circumstances as places or countries (Iraq, Israel, United States, etc.), gives the names of the interviewed persons or presents an important event (Olympic Games, hostage crisis, etc.). To extract text from stories shots we based on our work proposed in [4]. In this work we combine colour and edges to extract text. However, as the majority of approaches it includes three main tasks: detection, localization and segmentation. First step, we apply a new multiple frames integration (MFI) method to minimize the variation of the background of the video frames to eliminate from frame columns and rows of pixels which are not containing text. Second step we try to localize text pixels from the remaining rows and columns clusters. Every window is represented by two frames. One is the frame of the window filtered along rows and the other is the frame but filtered along the columns. For every frame we achieve two operations: First, we realize a transformation from the RGB space to HSV space. Second, we generate using Sobel filters, an edge picture.

For every cluster of these frames, we formulate a vector composed of ten features: five representing the HSV image and five representing the edge picture. These features are computed as follows: mean second order moment, third order moment, minimum value of the confidence interval and maximum value of the confidence interval.

The generated vectors will be presented to a trained back propagation neural network containing 10 inputs nodes, 3 hidden nodes and 1 output node .The training data base contains 2000 key frames with the dimension of 320x240. The results of the classifications are two images: an image containing rows considered as text rows and an image containing columns considered as text columns. Finally, we merge results of the two images to generate an image containing zones of text (Figure 4).

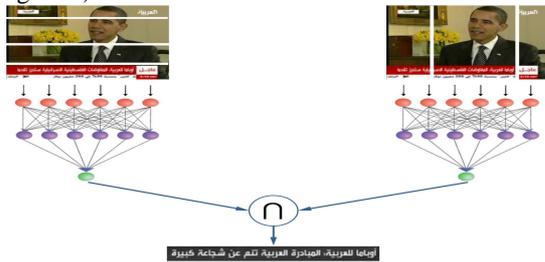

Figure 4: Text localization through neural networks

After localizing text in the frame, the following step consists in segmenting text. First, we compute the gray levels image. Second, for each pixel in the text area, we create a vector composed of two features: the standard deviation and the entropy of the 8 neighborhood of pixels. Third, we run the fuzzy C means clustering algorithm to classify the pixels into "text" cluster and "background" cluster. Finally, we binarize the text image by marking text pixels in black (Figure 5).

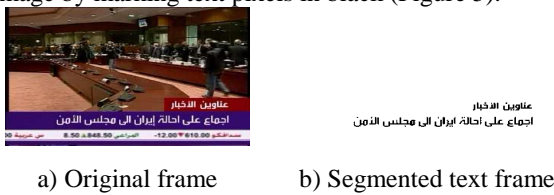

a) Original frame      b) Segmented text frame

Figure 5: Example of text extracted

## 4. Arabic Text Recognition

Text recognition, even from the detected text lines, remains a challenging problem due to the variety of fonts, colors, presence of complex backgrounds and the short length of the text strings. An optical character recognition methodology was implemented including five successive stages: Pre-processing, Segmentation, Feature extraction, Classification and Post-processing.

### 4.1. Structure of the Arabic Script

The alphabet from the Arabic language has 28 consonants including 15 from one to three points that differentiate between similar characters. The points and Hamza (ء) called secondary characters (complementary). They are located above the primary character as the "alif" (أ) below as the "Ba" (ب), or in the middle as the "jeem" (ج). There are four characters, which can take the secondary nature Hamzah (ء): alif (أ), waw (ؤ) kaf (ك) and ya (ئ) [5].

A distinctive feature of Arabic script is the existence of a base-line. The base-line is a horizontal line that runs through the connected portions of text. The baseline has the maximum number of text pixels (Figure 6).

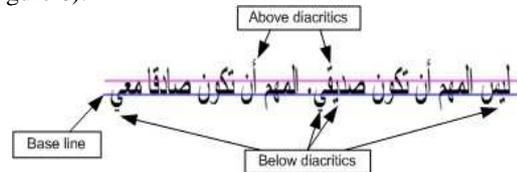

Figure 6: Arabic text showing

### 4.2. Pre-processing

The low resolution of video sequences is the major source of problems in text recognition. The pre-processing stage attempts to compensate for poor quality originals and/or poor quality of Binarization stage after text extraction. This is achieved by reducing both noise and data variations.

**4.2.1. Text Enhancement.** There are both the pixels of text and the pixels of background inside the located text block. So, a filtering step is necessary. For the filtering step, we use the Gabor Filter.

The Gabor filter $G_{\lambda,\theta,\varphi}(x,y)$ is used to obtain the spatial frequency of the local pattern in an image:

$$G_{\lambda,\theta,\varphi}(x,y) = e^{-\frac{(x'^2+\gamma^2 y'^2)}{2\sigma^2}} \cos(2\pi \frac{x'}{\lambda}+\varphi) \quad (1)$$

Where $x' = x\cos\theta + y\sin\theta$ and $y' = -x\sin\theta + y\cos\theta$

$x$ and $y$ represent the pixel coordinates, $\theta$ specify the orientation of the filter, $\gamma$ determines the spatial aspect ratio, and $\frac{1}{\lambda}$ is called the spatial frequency.

**4.2.2. Text Normalization.** In different images, text may occur with various widths and heights. To have consistent features through all the text images and to reduce the variance of the size of the characters, the height of a text image is normalized. The height of the text image is scaled to a predefined size (26 pixels).

### 4.3. Text segmentation

All components that do not have pixels belonging to the baseline are considered diacritic. The baseline is the one that corresponds to the maximum of the horizontal projection. Information concerning the nature and location of these diacritical marks are registered for use in the future at the stage of recognition.

After baseline detection and diacritic elimination, a segmentation stage is necessary. This segmentation is based on a vertical projection. The segmenter should avoid cases of under-segmentation and/or over-segmentation. Each segment will be recognized using a base of segments.

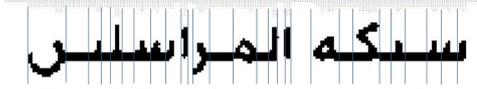

Figure 7: Example of text segmentation

### 4.4. Feature extraction

Feature extraction is one of the most difficult and important problems of pattern recognition. The feature extraction step is closely related to classification, because the type of features extracted here must match what the classifier expects. The Features chosen to represent each character are:

- Extraction of occlusion: They correspond to the internal contour of the primary plots characters.
- The projection Feature: In addition to the horizontal and vertical projection, we make a projection of the diagonal and the slanting diagonal.
- Extraction of diacritics marks: diacritics are the secondary parts of characters. These marks above and below the characters, have an important role in the distinction of a few characters that differ only by the number or location of points. The following characters (ب ت ث ي ن) are differed only by the number and location of points.
- The transition Features: the number of transition from 0 to 1 of the row, column, diagonal and slanting diagonal. The result is a vector of 160 features for each image of a letter

### 4.5. Classification

Classification in an OCR system is the main decision making stage in which the extracted features of a test set are compared to those of the model set. Based on the features extracted from a pattern, classification attempts to identify the pattern as a member of a certain class. When classifying a pattern, classification often produces a set of hypothesised solutions instead of generating a unique solution.

Supervised classification methods can be used to identify a sample pattern. We use the fuzzy k-nearest neighbour algorithm (k=10). The coefficient of belonging of a new segment $x_i$ to class j is given by the formula:

$$u_{ij} = \frac{\sum_{t=1}^{k} u_{jt} \left( \|x_i - x_t\|^{m-1} \Big/ 2 \right)}{\sum_{t=1}^{k} \left( \frac{1}{\|x_i - x_t\|^{m-1} \Big/ 2} \right)} \quad (2)$$

Where $u_{ij}$ is the coefficient of belonging to class $w_j$, the $t^{th}$ observation, among the k nearest neighbors of $x_i$. M is the variable determines the importance of the contribution of the distance in the calculating of the function of belonging.

### 4.6. Post-processing

The last step is the post-processing. It improves the recognition by refining decisions previously prized in the classification stage and recognises words by using context. It is ultimately responsible for issuing the best solution and is often implemented as a set of techniques, based on character frequencies, lexicons and other context information.

## 5. Experiments

To validate our approach, we have used a varied database composed of news sequences extracted from different Arabic channels. Concerned channels are TV7 Tunisia *(tunisiatv.com)* as generalist channel presenting news at 13H and 20H. We have also tried our approach with news video extracted from Aljazeera *(Aljazeera.net)* and Al Arabia *(alarabiya.net)* which are specialist channels presenting news continually. We have used the following measures of precision and recall to evaluate the segmentation method.

$$recall = \frac{A_{find}}{A_{total}} = \frac{A_{find}}{A_{find} + A_{miss}} \quad (3)$$

$$precision = \frac{A_{find}}{A_{find} + A_{false}} \quad (4)$$

$A_{find}$: the number of the right detected stories
$A_{false}$: the number of the false detected stories
$A_{miss}$: the number of neglected stories

| Channel | Duration | Recall | Precision |
|---|---|---|---|
| TV7 Tunisia | 10 hours | 89.66% | 88% |
| Al Jazeera | 10 hours | 90.53% | 93.3% |
| Al Arabia | 10 hours | 93.45% | 91.22% |

Table 1. Evaluation results of textual localization

We notice that Al Jazzeera and Al Arabia channels present the best rate of recall and precision because they present the best quality of graphic text.

To evaluate our framework of text extraction we have based on the number of identified correct text contours, we find the following recall and precision rate:

|  | Channel | Duration | Recall | Precision |
|---|---|---|---|---|
| **Our approach** | TV7 Tunisia | 10 hours | 89.66 | 88 |
|  | Al Jazeera | 10 hours | 90.53 | 93.3 |
|  | Al Arabia | 10 hours | 93.45 | 91.22 |
| **L. Agnihotri and al. [11]** | TV7 Tunisia | 10 hours | 86.43 | 84,35 |
|  | Al Jazeera | 10 hours | 87.51 | 90.1 |
|  | Al Arabia | 10 hours | 90.15 | 87.22 |

Table 2. Evaluation results of text localization

In order to investigate the effectiveness of our recognition sub-system in recognizing Arabic Text extracted from various news sequences, a series of tests were performed using fuzzy KNN classifier with K=3, K=5 and K=10. Table 2 shows the recognition results. The word and character recognition rate are defined as follows:

$$CR(WR) = \frac{\text{Correctly Recognized Characters (Words)}}{\text{Total Number of Characters (Words)}} \times 100 \quad (3)$$

| Channel | K=3 | K=5 | K=10 |
|---|---|---|---|
| TV7 Tunisia | 80% | 82% | 85% |
| Al Jazeera TV | 83% | 87% | 92% |
| Al Arabia TV | 83% | 90% | 95% |

Table 3: Evaluation results of text recognition

We note that the rate of recognition of texts extracted from Al-Arabiya and Al-Jazeera TVs are better than those extracted from Tunis 7 TV. This is explained by the fact that the text extract from Al-Arabiya and Al-Jazeera TVs is clearer and more readable. We find it is difficult to recognize small size characters that appear frequently in news videos. The best recognition rate is in all cases for K=10.

The recognition system is tested on the Data Base of characters provided by ICDAR 2009 Handwritten Farsi/Arabic Character Recognition Competition, under the name REGIM [15]. The average recognition rate of characters is almost 95%.

## 6. Conclusion

In this paper, we propose an efficient method to deal with background complexity in Arabic news video text recognition by efficiently integrating multiple frame information. By using this method we can produce quite clear text for Arabic text recognition. The extraction rate has been increased about 91% and 84% for text recognition. These methods can also be adopted by any other type of Video OCR systems to increase recognition rate.

## Acknowledgement

The authors would like to acknowledge the financial support of this work by grants from General Direction of Scientific Research (DGRST), Tunisia, under the ARUB program.